\documentclass[12pt]{article}
\usepackage{epsfig}
\usepackage{a4,amsmath}
\begin{document}
\title{Theoretical challenges in atom optics:\\
 Atomic and molecular
  diffraction by transmission gratings\footnote{Invited plenary
    lecture at the International Symposium {\em New Insights
in Quantum Mechanics}, Goslar, September 1998. Proceedings edited by
H.-D. Doebner, S.T. Ali, M. Keyl and R.F. Werner, World Scientific
1999.
}}
\author{Gerhard C. Hegerfeldt and Thorsten K\"ohler\\
Institut f\"ur Theoretische Physik, Bunsenstrasse 9,\\ D-37073
  G\"ottingen, Germany}
\date{}
\maketitle
\begin{abstract}
A few years ago, diffraction of atoms by double slits and gratings was
achieved for the first time, and standard optical wave-theory provided an
excellent description of the experiments. More recently, diffraction
of weakly bound molecules and even clusters has been observed. Due to
their size and to possible breakup processes optical wave-theory is no
longer adequate and a more sophisticated approach is needed. Moreover,
surface effects, which can modify the diffraction pattern of atoms and
molecules, give rise to further complications. In this article a fully
quantum mechanical approach to these questions is discussed.
\end{abstract}

\section{Introduction}
New and interesting theoretical challenges are presently arising in atom and
molecular optics, as we want to show here.
The rapidly growing field of atom optics exploits wave-theoretical aspects of
quantum-me\-chan\-ical particles and their interference effects. For
electrons and neutrons this has a long history, but for atoms it is
more recent. Advances in nanotechnology have made it possible to
manufacture structures such as double slits and gratings with slit
widths as small as 50 nm. Typical optics experiments have been carried
over to atoms. For example, atomic beams have been sent through a
double slit and diffraction patterns were observed 
\cite{Mlynek}. Similar
experiments have been performed with transmission gratings and large
numbers of diffraction peaks were observed with high resolution 
\cite{Pritchard,SchoellToenn}.

In such experiments the incoming atoms usually can be described by
plane waves which are associated with quantum-mechanical point
particles. Classical optics with its standard wave-theoretical methods
and approximations, in particular those of Huygens and Kirchhoff, has
been successfully applied to atom optics and has yielded good
agreement with experiments.

Interesting problems with 
this simple picture have, however, appeared 
quite recently. One
complication arises in the diffraction of molecules, in particular of
weakly bound systems. When a diffracted system is observed at a
nonzero angle from the incident direction, then the system has
received a momentum transfer from the grating. For a weakly bound
system this may induce breakup processes and result in a change of the
diffraction pattern. Moreover, weakly bound systems are large 
compared to
atoms, and treatment as point particles may no longer be
adequate. In addition, low-lying molecular rotational or vibrational
states may be excited by the passage through the grating, and this may
also change the diffraction pattern. As a further complication,
surface potentials resulting from van der Waals forces have been seen
to change diffraction patterns \cite{Grisentietal}.

Experimentally, molecular diffraction has been observed for He$_2$ in
Ref.~\cite{SchoellToenn} and for Na$_2$ in 
Ref.~\cite{PritchardNa2}. The molecule Na$_2$ is tightly
bound and very small 
compared
to present-day slit widths, and the results seem to
be describable 
by classical wave-optics. The recently discovered
helium dimer He$_2$ 
\cite{SchoellToenn,Gentrydimer}, however, is extremely weakly bound 
($-$0.11 $\mu$eV) and quite large, 
with an estimated diameter of about
6 nm \cite{Gentry}. For a slit width of 50 nm this considerable size of the
molecule has some effect on the diffraction pattern, and this effect
will increase for future smaller slit widths.

We have performed a theoretical analysis of diffraction of weakly
bound diatomic systems by a transmission grating \cite{HeKoe}. The main
point which we wish to stress in this article is that 
{\em diffraction} by a transmission grating really means {\em scattering}
by the grating bars. Therefore, since breakup processes and finite-size
effects are to be included, the proper approach is to consider quantum
mechanical scattering of composite systems by an external potential,
$W$ say. For the beam velocities used in experiments the electronic
degrees of the two atoms can be neglected, and thus one deals with a
system of two particles, bound by a potential, denoted by $V$. 
Due to the presence of the external potential $W$ resulting
from the grating bars ("walls") the
problem is quite different from one-particle scattering and rather
resembles, at least mathematically, the three-body problem where,
loosely speaking, the external potential plays the role of a third body. 

Not surprisingly one can therefore adapt methods of few-body theory to
the present problem, in particular the Alt-Grassberger-Sandhas (AGS)
equations \cite{AGS} 
which are related to the Faddeev equations. In Section 2
we briefly outline how one can adapt this to the problem at hand, and we
obtain AGS equations for the scattering amplitude of a bound
two-particle system by an external potential. These equations
decouple here and can be iterated in a straightforward way. In
Section 3 we indicate how the lowest order can be specialized to
diffraction scattering by a transmission grating with a short-range 
strongly repulsive ("reflective") bar potential. Our results are
illustrated for He$_2$. 
At the end we will discuss the role of weakly
attractive surface potentials and how they can be incorporated.

\section{Scattering of a composite 
two-particle system by an external potential}
In this section the quantum mechanical scattering problem of a bound 
two-particle system (molecule) by a quite general obstacle will be 
considered. The latter is described by an external potential 
of the form
\begin{equation} \label{W}
  W({\bf x}_1,{\bf x}_2)=W_1({\bf x}_1)+W_2({\bf x}_2)
\end{equation} 
where ${\bf x}_{1,2}$ is the position of the respective 
atomic constituent and $W_{1,2}$ the respective single-atom potential.
The binding potential is denoted by $V({\bf x}_1-{\bf x}_2)$, 
with negative binding energies 
$E_\gamma$ and two-particle bound states 
$\phi_\gamma({\bf x}_1-{\bf x}_2)$, 
where the potentials $V$, $W_1$ and $W_2$ are 
assumed to be of short range.
Among all possible scattering channels
in the following only those will be considered in which the molecule 
is elastically scattered or changes its internal bound state. 
The plane-wave scattering matrix element for an incoming molecule 
with momentum ${\bf P}'$ and 
internal state $\phi_{\gamma'}$ and 
an outgoing molecule with momentum ${\bf P}$ 
and internal state $\phi_\gamma$ reads
\cite{Newton}
\begin{eqnarray}
  \label{Smatrix}
  \langle{\bf P},\phi_\gamma|S|{\bf P}',\phi_{\gamma'}\rangle
  &=&\delta_{\gamma\gamma'}\delta^{(3)}({\bf P}-{\bf P}')\\
  \nonumber
  &&
  -2\pi i\delta
  \left(
    \frac{|{\bf P}|^2}{2M}+E_\gamma-\frac{|{\bf P'}|^2}{2M}-E_{\gamma'}
  \right)
  t({\bf P},\phi_\gamma;{\bf P}',\phi_{\gamma'})
\end{eqnarray}
where $M$ is the molecular mass.
The transition amplitudes 
$t({\bf P},\phi_\gamma;{\bf P}',\phi_{\gamma'})$
determine the molecular diffraction pattern and can be calculated by
adapting the AGS 
\cite{AGS}
approach as follows.
Introducing the Green's operators 
\begin{eqnarray}
  G_0(z)&\equiv&(z-H_0)^{-1},\\
  \nonumber
  G(z)&\equiv&(z-H_0-V-W)^{-1},\\
  \nonumber
  G_V(z)&\equiv&(z-H_0-V)^{-1},\\
  \nonumber
  G_W(z)&\equiv&(z-H_0-W)^{-1},
\end{eqnarray}
where $H_0$ denotes the sixdimensional
free Hamiltonian (kinetic energy) of two particles,
the operators $U_{VV}(z)$ and $U_{WV}(z)$ are defined by
the relations
\begin{align}
  \label{transopUVV}
  G(z)\equiv G_V(z)+&G_V(z)U_{VV}(z)G_V(z),\\
  \label{transopUWV}
  G(z)\equiv \quad\quad\quad\ \ &G_W(z)U_{WV}(z)G_V(z).
\end{align}
As in Ref.~\cite{AGS}
it can be shown that $U_{VV}$ is directly related to the scattering
matrix Eq.~(\ref{Smatrix}) through
\begin{equation}
  \label{transamptransop}
  t({\bf P},\phi_\gamma;{\bf P}',\phi_{\gamma'})=
  \langle{\bf P},\phi_\gamma|U_{VV}
  \left(
    |{\bf P}|^2/2M+E_\gamma
    +i0
  \right)
  |{\bf P}',\phi_{\gamma'}\rangle.
\end{equation}
The resolvent equation 
$G(z)=G_V(z)+G_V(z)WG(z)$,
applied to the left hand side of Eq.~(\ref{transopUVV}), gives
\begin{equation}
  \label{AGS1}
  G_V(z)+G_V(z)WG(z)=G_V(z)+G_V(z)U_{VV}(z)G_V(z),
\end{equation}
and Eq.~(\ref{transopUWV}) inserted into the left-hand side of
Eq.~(\ref{AGS1})
gives
\begin{equation}
  \label{AGS2}
  U_{VV}(z)=WG_W(z)U_{WV}(z).
\end{equation}
We introduce the $T$ matrix 
$T_W(z)\equiv W+WG_W(z)W$,
which satisfies
\begin{equation}
  T_W(z)G_0(z)=WG_W(z)
  \left(
     G_W^{-1}(z)+W
  \right)
  G_0(z)=WG_W(z).
\end{equation}
Eq.~(\ref{AGS2}) can then be written as 
\begin{equation}
  \label{AGSI}
  U_{VV}(z)=T_W(z)G_0(z)U_{WV}(z).
\end{equation}
In a similar way a second AGS-like equation
is obtained as 
\begin{equation}
  \label{AGSII}
  U_{WV}(z)=G_0^{-1}(z)+T_V(z)G_0(z)U_{VV}(z),
\end{equation}
where $T_V\equiv V+VG_V V$, in analogy to $T_W$.
Once iterated Eqs.~(\ref{AGSI}) and 
(\ref{AGSII}) decouple and give 
\begin{equation}
  \label{AGSresult}
  U_{VV}(z)=T_W(z)+T_W(z)G_0(z)T_V(z)G_0(z)U_{VV}(z).
\end{equation}
Although Eq.~(\ref{AGSresult}) explicitly determines only 
the transition 
operator for elastic and excitation processes, $U_{VV}(z)$, all 
channels are implicitly included. Similar to 
Eq.~(\ref{transamptransop})
it can further be shown that 
\begin{equation}
  U_{0V}(z)=G_0^{-1}(z)+(1+T_V(z)G_0(z))U_{VV}(z)
\end{equation}
is the transition operator for a molecular break-up into
two asymptotically free particles.

Eq.~(\ref{AGSresult}) is exact and contains both breakup and finite-size
effects. It lends itself to iteration, and to lowest order in $T_V$ one has 
\begin{equation}
  U_{VV}\cong T_W.
\end{equation}
By Eq.~(\ref{transamptransop}) this gives
\begin{equation}
  \label{21}
  t({\bf P},\phi_\gamma ;{\bf P}^\prime,\phi_{\gamma^\prime})
  \cong \langle {\bf P},\phi_\gamma |T_W(E+{\rm i}0)
  |{\bf P}^\prime,\phi_{\gamma^\prime}\rangle
\end{equation}
with $E=|{\bf P}|^2/2M+E_\gamma$.
Although $T_W$ is the transition operator for scattering of two 
asymptotically free
particles by the external potential $W$ it is not on the energy shell
in Eq.~(\ref{21}). The interaction $V$ is taken into account 
in Eq.~(\ref{21}) through
the wave-functions $\phi_\gamma$ and $\phi_{\gamma'}$.

\section{Application to diffraction  gratings}
Although the potential $W = W_1 + W_2$ is additive in the two
particles and looks very simple, Eq.~(\ref{21}) 
is not straightforward
to evaluate since the Green's operator $G_W$ is not easily expressed
by one-particle Green's operators. In the diffraction domain the de
Broglie wavelength is small compared to the extension of the obstacle
and the potential. To evaluate Eq.~(\ref{21}) 
for small scattering
angles one obtains an integral expansion of 
Eq.~(\ref{21}) by
inserting two complete sets of two-particle plane waves, 
$|\mathbf p_1, \mathbf p_2 \rangle$ and 
$|\mathbf p_1', \mathbf p_2'\rangle$, on the
right-hand-side. The kinetic energies of these states are not equal to
$E$ (i.e. $T_W$ is not on the energy shell). However, 
the momentum space wave-functions
$\phi_\gamma(\mathbf p)$ 
and $\phi_{\gamma'} (\mathbf p)$ are centered at relative momenta
${\bf p}$ which are 
small compared to $P\equiv |{\bf P}|$, 
and this fact can be used to show 
that in $T_W(E+i0)$ one can replace $E$ by 
$\mathbf p^2_1/2m + \mathbf p^2_2/2m$, where, for simplicity, 
the atomic constituents
are assumed to have the same mass, $m=M/2$. 
The Lippmann-Schwinger equation for the potential $W$
further enables one to reduce Eq.~(\ref{21}) to one-particle
transition amplitudes for the potentials $W_1$ and $W_2$. For
reflecting bars these can be calculated by standard methods.

\begin{figure}[hbt]
\begin{center}
\epsfig{file=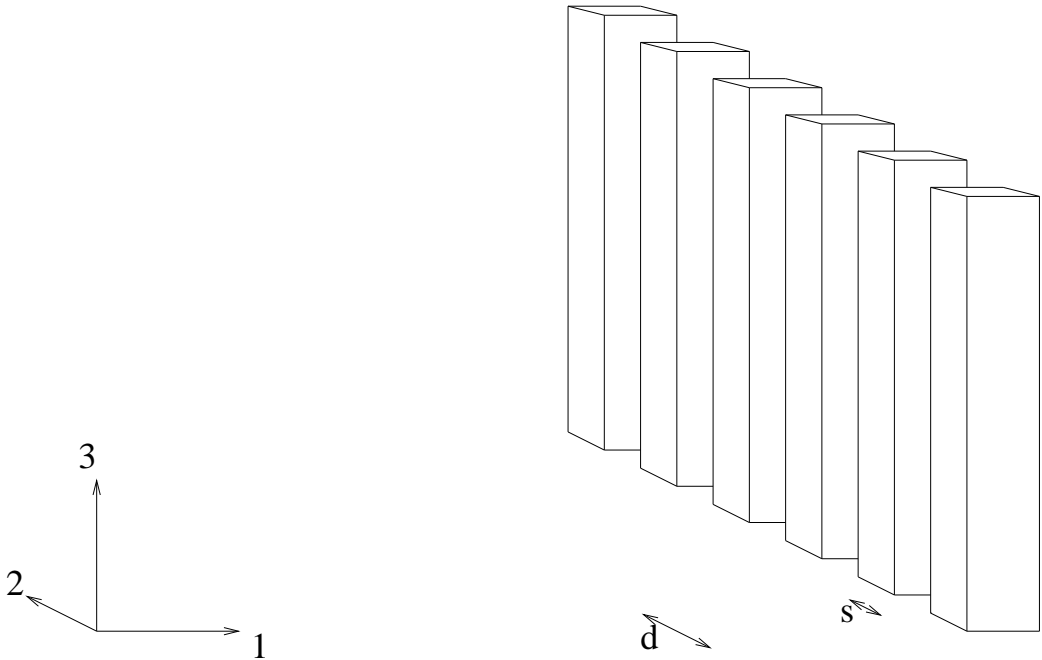,width=250pt,angle=270}
\end{center}
\caption[]{Transmission grating with slit width $s$ and grating period
  $d$.\label{fig:grating}} 
\end{figure}

If the bar and slit widths are much larger than the diameter of the
two-particle system, then the one-particle amplitudes are sharply
peaked at the forward direction on a scale on which the other
expressions appearing in the integral expansion of 
Eq. (\ref{21}) are
slowly varying and can therefore be dropped. This then leads to the
usual diffraction pattern of a point particle of mass 2$m$.

When bar and slit widths decrease the situation changes. In this case
one obtains a more complicated coherent amplitude part, $t_{coh}$, the
superposition of amplitudes for the interaction of the molecule with
individual bars. The incoherent part from the interaction with several
bars becomes appreciable only when $d$ and $s$ 
(see Fig.~\ref{fig:grating}) are of the order of the
molecular diameter. The net result is of the form
\begin{equation}
  \label{39}
  t_{\rm coh}(P_2)=t^{\rm mol}_{\rm bar}(\gamma,P/M;P_2)
    \frac{\sin(P_2 N d/2\hbar)}{\sin(P_2 d/2\hbar)}  \delta(P_3)
\end{equation}
where the first factor is the molecular transition amplitude for a
single bar of width $d-s$, for total momentum $P$ and total mass
$M$.  The second factor is the usual sharply
peaked grating function known from optics, and it gives the
diffraction peaks at the same locations as for
point-particles with corresponding lateral momentum transfer. The
$\delta$ function is due to the assumed infinite extent of the grating
in the third direction. The amplitude of
the diffraction  peaks is determined by $t^{\rm mol}_{\rm bar}$. The
latter has been 
explicitly calculated by means of Eq.~(\ref{21}) 
in Ref.~\cite{HeKoe}
and it depends only on the absolute value of the ground-state
wave-function. One obtains 
\begin{eqnarray}
  \label{40}
&\!\!\!\!\!\!\!\!\!\!\!\!\!\!\!\!\!\!\!\!\!\!\!\!\!\!\!\! 
t^{\rm mol}_{\rm bar}(\gamma,P/M;P_2)
 =-\frac{2{\rm i} P}{(2\pi)^2 M}
  \frac{\sin[P_2(d-s)/2\hbar]}{P_2}
  \int{\rm d}^3 x
      \,2\,{\rm e}^{{\rm i} P_2 x_2/2 \hbar}
  |\phi_\gamma({\bf x})|^2\\
  \nonumber
  &~~~~~~~~~~~~~~~+\frac{4 {\rm i} P}{(2\pi)^2 M}
  \int{\rm d} x_1 {\rm d} x_3 
  \int_0^{d-s} {\rm d} x_2
  |\phi_\gamma({\bf x})|^2
    \sin
    \left[
      \frac{P_2}{2\hbar}(d-s-x_2)
    \right]
      /P_2~.
\end{eqnarray}
The expression preceding the first integral 
is the single-bar amplitude for a
point particle of mass M and total momentum $P$. 
The complete expression on the
r.h.s. reduces to this if 
$|\phi_\gamma({\bf x})|^2$ contracts to a point.

\begin{figure}[hbt]
\begin{center}
\epsfig{file=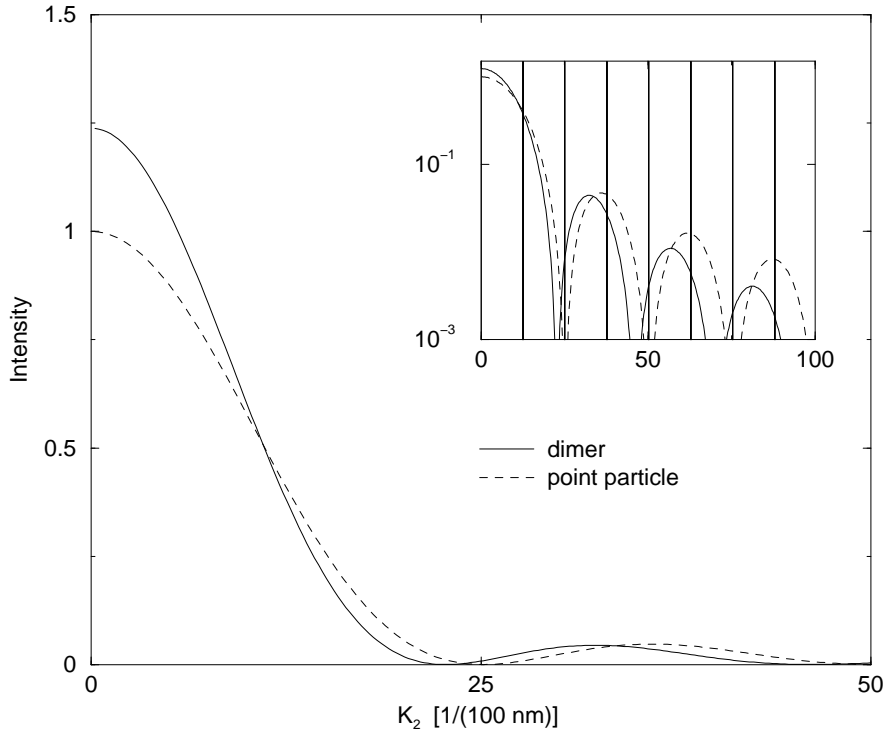,width=280pt,angle=270}
\end{center}
\caption[]{Diffraction by a single bar of width 25 nm for helium dimers and 
   point particles. For a symmetric grating with this bar and slit width
   the vertical lines in the inset pick out the corresponding
   diffraction peaks. For dimers there are even order peaks which
   are absent for point particles. The lateral momentum transfer is
   $P_2 =\hbar K_2$.\label{fig:a25molwire}} 
\end{figure}

In Fig.~\ref{fig:a25molwire} 
we have plotted $|t^{\rm mol}_{\rm bar}|^2$ for He$_2$ and the
corresponding quantity for a point particle. The bar width is 25 nm.
For a symmetric grating of period 50 nm the vertical lines in the inset 
indicate where the peaks of the grating functions cut out the diffraction
peaks of first order, second order and so on. For a symmetric grating
and point particles there are no 
even-order diffraction peaks since the single-bar amplitude vanishes
there, as  seen in  the inset. For the dimer this is not true, and it 
 therefore has small even-order peaks which become more pronounced for
 smaller bar and slit widths.

\begin{figure}[hbt]
\begin{center}
\epsfig{file=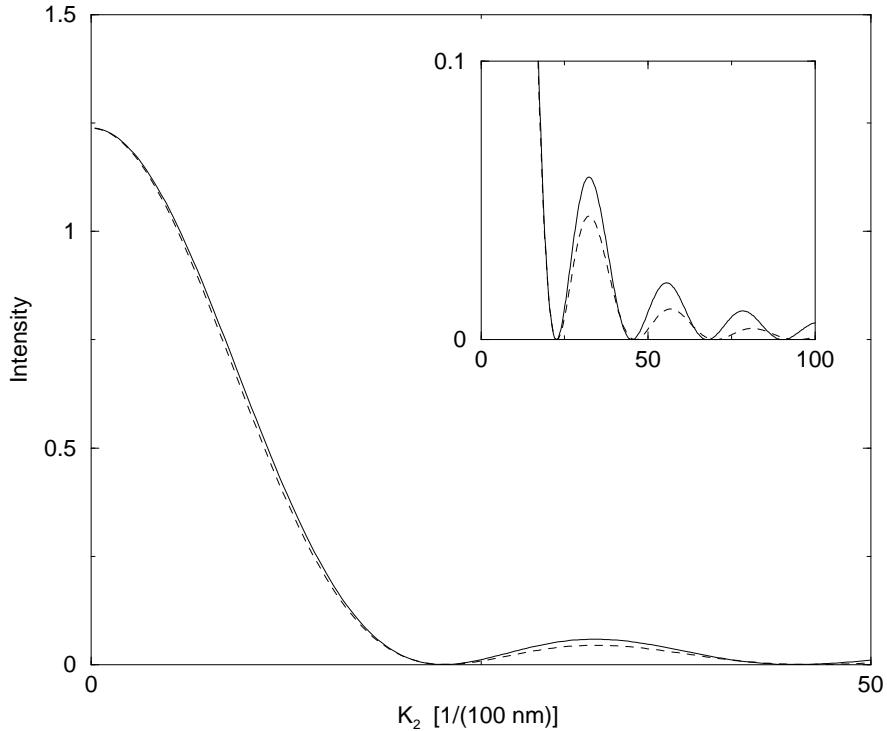,width=280pt,angle=270}
\end{center}
\caption[]{Diffraction of helium dimers (solid line) by a single bar
  of width 25 nm and diffraction of point particles (dashed line) by a
  single bar of width (25 + 2.8) nm. There is agreement for small
  lateral momentum transfer $P_2 =\hbar K_2$.
  \label{fig:a25ekum}} 
\end{figure}

Intuitively we expect that finite-size effects of He$_2$ might be
partially taken into account by a larger bar and smaller slit
width. Fig.~\ref{fig:a25ekum} 
shows that the single-bar amplitude of He$_2$, for a bar
width of 25 nm, can be approximated for small lateral momentum transfer by
that of a point particle for bar width of (25+2.8) nm. For a grating
this would mean a nonsymmetric grating of correspondingly smaller slit
width. The expectation value of $|x_2|$ for the ground-state
wave-function of He$_2$ we are using is just 2.8 nm. 

\section{The role of attractive surface potentials}
When a rare-gas atom such as helium passes through a slit it
experiences a repulsive potential of very short range. 
However, if the atom is
polarizable there will also be a longer-range attractive surface
potential acting on it, due to van der Waals forces. Recent
experiments  
\cite{Grisentietal}
have shown that this additional attractive potential
can have an unexpectedly large influence on the diffraction pattern,
not so much so 
for helium but rather for the heavier rare-gas atoms, 
which have a higher polarizability.

The attractive part of the surface potential, denoted by $W_S$, is of
the form
\begin{equation*}
W_S=-\frac{C_3}{r_\perp^3}
\end{equation*}
where $r_\perp$ is the distance of the atom from the slit surface and
where $C_3$ is a constant depending on the grating material and the
particular rare-gas atom.

\begin{figure}[hbt]
\begin{center}
\epsfig{file=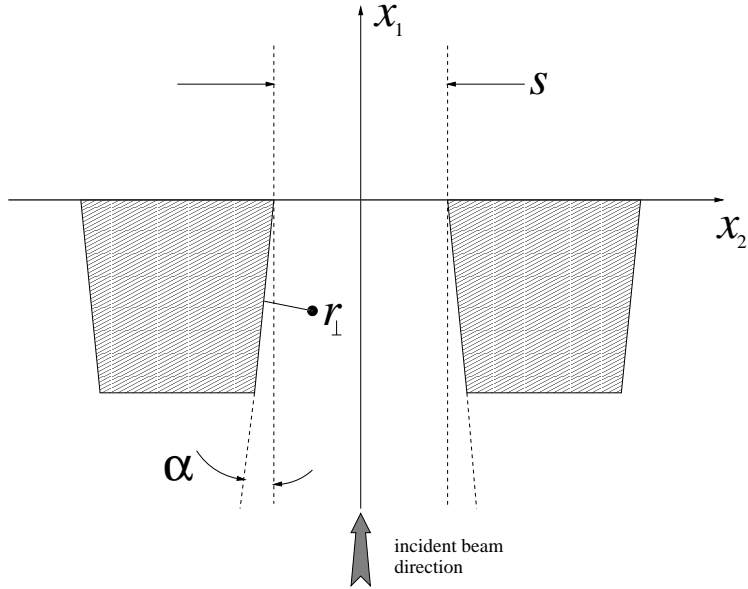,width=280pt,angle=0}
\end{center}
\caption[]{Trapeze-like cross section of grating bars.
\label{fig:wedge}} 
\end{figure}

In a first step we have taken $W_S$ into account for the atomic case,
with quite an unexpected and surprising result. The simple picture of
the grating given in Fig.~\ref{fig:grating} 
as an array of rectangular bars is not
adequate if the surface potential plays a role. In fact, with the
simple geometry of Fig.~\ref{fig:grating} 
our results did not even qualitatively agree
with early experiments for heavier rare-gas atoms. 
As it then turned out, the
etching process produces bars whose cross section is not quite
rectangular but rather trapeze-like as in Fig.~\ref{fig:wedge}, 
with a small angle
$\alpha$ of about $8^\circ$ off the perpendicular direction. This
seems to be a small deviation from the idealized rectangular form of
Fig.~\ref{fig:grating}, 
but taking the angle $\alpha$ into account gave at least
qualitative agreement. In the mean time the approach, both
experimentally and theoretically, has been so perfected, that it has
been possible to determine both the grating parameters as well as the
surface potential constant $C_3$ from experimental data 
\cite{Grisentietal}.

For dimers the surface potential can be taken into account by adding
$W_S^{(1)}$ and $W_S^{(2)}$ to $W$ in Eq.~(\ref{W}). The details are
presently being worked out.

\section{Conclusions}
We have shown in this article that 
new and interesting theoretical questions arise in atom optics when the
spatial extent of a diffracted 
system can no longer be ignored and when possible
excitation energies become comparable to the kinetic energy of
diffracted systems. Here we have studied the diffraction of a weakly
bound two-particle system by an obstacle, in particular by a
transmission grating. We have obtained an analytic expression for the
elastic diffraction amplitude which incorporates breakup processes and
finite-size effects. This expression contains the familiar grating
function of classical optics as a factor and this puts the diffraction
peaks at the angles expected from classical wave theory. This grating
function is multiplied by the molecular diffraction amplitude of a
single grating bar, and in this factor a difference from the
point-particle case may arise.

For a molecular diameter very small compared to the bar and slit
widths the derivations from the point-particle result are indeed
negligible, as expected. This is no longer true for smaller bar and
slit widths. In particular, heights of diffraction peaks may become
lower than for point-particles. This can be attributed to breakups
that diminish the actual number of molecules in a given
direction. Moreover, the decrease in peak height goes with the
diffraction angle and thus with the lateral momentum transfer, which
is quite reasonable since higher momentum transfer should imply more
breakups. In addition, diffraction orders that are suppressed to zero
for point particles and for symmetric gratings may reappear.

If the slits become smaller than the diameter of the two-particle
system, additional terms appear in the diffraction patterns. But this
case is presently not of experimental interest.

The appearance of the diffraction peaks at the wave-theoretical
locations for gratings holds quite generally, also for larger
composite systems. For dimers a simplified argument has been given by
us in Ref. \cite{fewbody}, and the general case will be treated
elsewhere.  

\end{document}